\documentclass[11pt]{article}
\usepackage{moriond,epsfig,caption2}

\def\be{\begin{equation}}
\def\ee{\end{equation}}
\def\bea{\begin{eqnarray}}
\def\eea{\end{eqnarray}}

\newcommand{\met}       {\mbox{$\not\!\!E_T$}}

\newcommand{\rargap}    {\mbox{ $\rightarrow$ }}
\newcommand{\ppbar}     {\mbox{$p\bar{p}$}}
\newcommand{\tbar}      {\mbox{$\bar{t}$}}
\newcommand{\ttbar}     {\mbox{$t\bar{t}$}}

\begin{document}
\vspace*{4cm}
\title{Top Quark Production at the Tevatron}

\author{ Liang Li }

\address{Department of Physics and Astronomy, University of California, Riverside, CA 92521, USA}

\maketitle\abstracts{
Top quark physics has been a rich testing ground for the standard model since the top quark discovery
in 1995. The large mass of top quark suggests that it could play a special role in searches 
for new phenomena. In this paper I provide an overview of recent top quark production cross section measurements 
from both CDF and D0 collaborations and also some new physics searches done in the top quark sector.
}

\section{Introduction}
Top quarks are produced in pair via the strong interactions or singly via the electroweak interactions
at hadron colliders. The top quark pair production gives a larger yield~\cite{ttbar-xsec} and provides more 
discrimination against backgrounds compared to the single top quark production. This is the main reason why the former was
first discovered in 1995~\cite{top-obs-1995-cdf,top-obs-1995-d0} and only after 14 years the later was observed at
the Tevatron Collider~\cite{stop-obs-2009-d0,stop-obs-2009-cdf}. 
Due to the large mass of the top quark, many models of physics beyond the standard model (BSM) predict observable 
effects in the top quark production rate. Measurements of top quark production cross section
serve as tests of possible new physics processes and can place stringent limits on these models.
In the standard model (SM), top quarks decay almost
100\% of the time to a $W$ boson and a bottom quark. The signature of top quark events is therefore defined
by the decay products of the $W$ boson. For $\ttbar$ events, if two $W$ bosons decay leptonically and there are 
two leptons in the final state, it is defined as the ``dilepton" channel of the $\ttbar$ production. Similarly, 
the ``all-hadronic" channel is defined when both $W$ bosons decay hadronically and the ``lepton+jets" channel is defined
when one $W$ boson decays leptonically and the other decays hadronically. The all-hadronic channel has the largest
branching ratio (BR) however also the lowest signal-to-background (S:B) ratio due to high multijets background. 
The dilepton channel has the highest S:B ratio however the signal statistics is limited by the lowest BR. Thus the 
most precise measurements on the top quark pair production rates are obtained in the lepton+jets channel. In the case of 
single top quark production, the cross section measurement is also done using the lepton+jets channel.

\section{Top Quark Pair Production}
\subsection{Lepton+jets channel}
In this channel, the $\ttbar$ events are identified using the decay of one $W$ boson to quarks and the other to a 
lepton and a neutrino. Each event is required to have a single high-$p_T$ electron or muon (for taus, only leptonically
decaying taus are considered) and at least three reconstructed jets. To suppress the background processes, at least one 
identified $b$-jet is required using the lifetime-based $b$-tagging algorithm~\cite{btagging}. The dominant background
is the $W$+jet production and other backgrounds are $Z$+jet, diboson ($WW$, $WZ$, $ZZ$), single top quark and multijet
processes.

\begin{figure}[!h!tbp]
\centering
\includegraphics[width=0.48\textwidth]{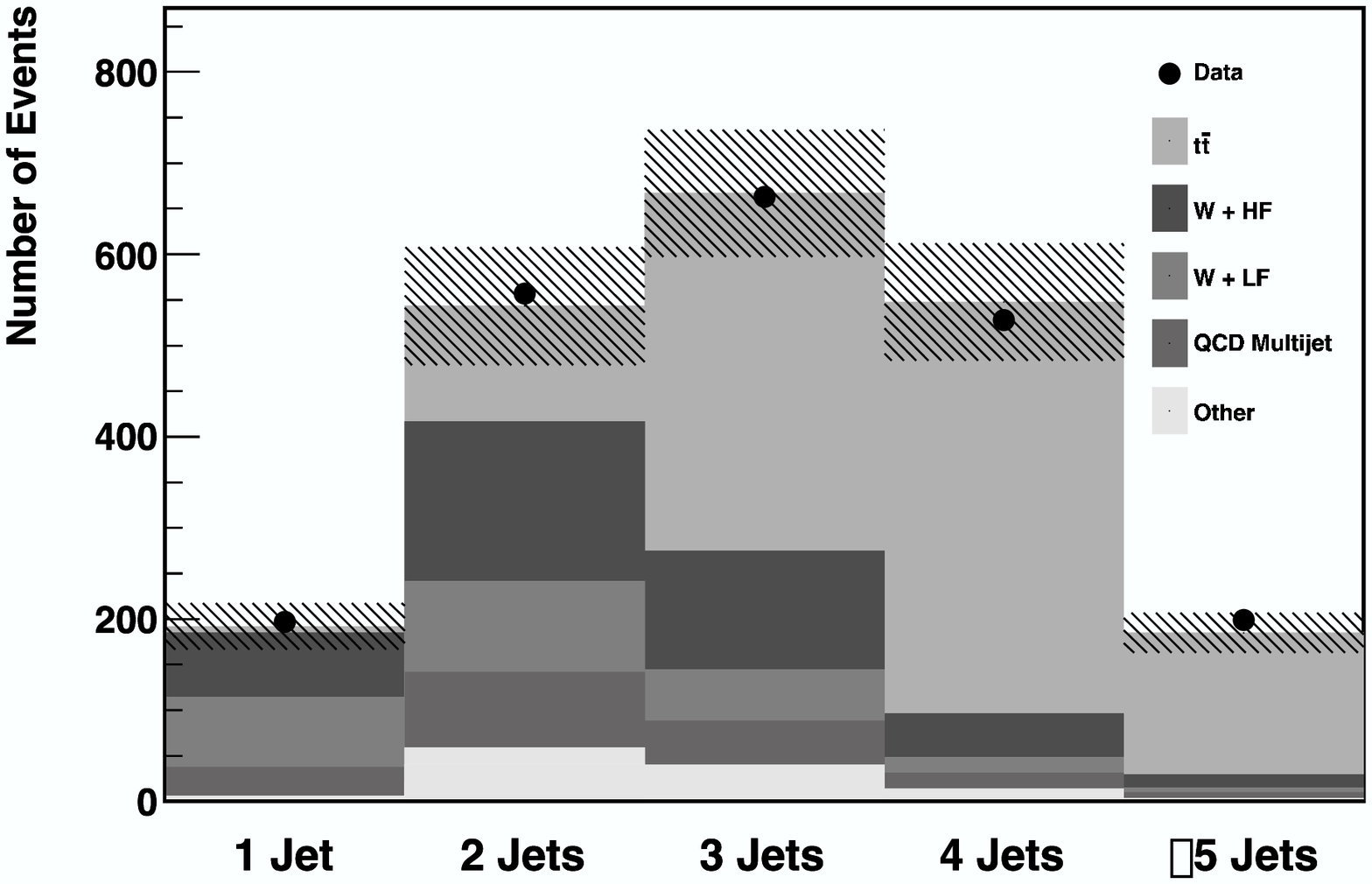}
\includegraphics[width=0.48\textwidth]{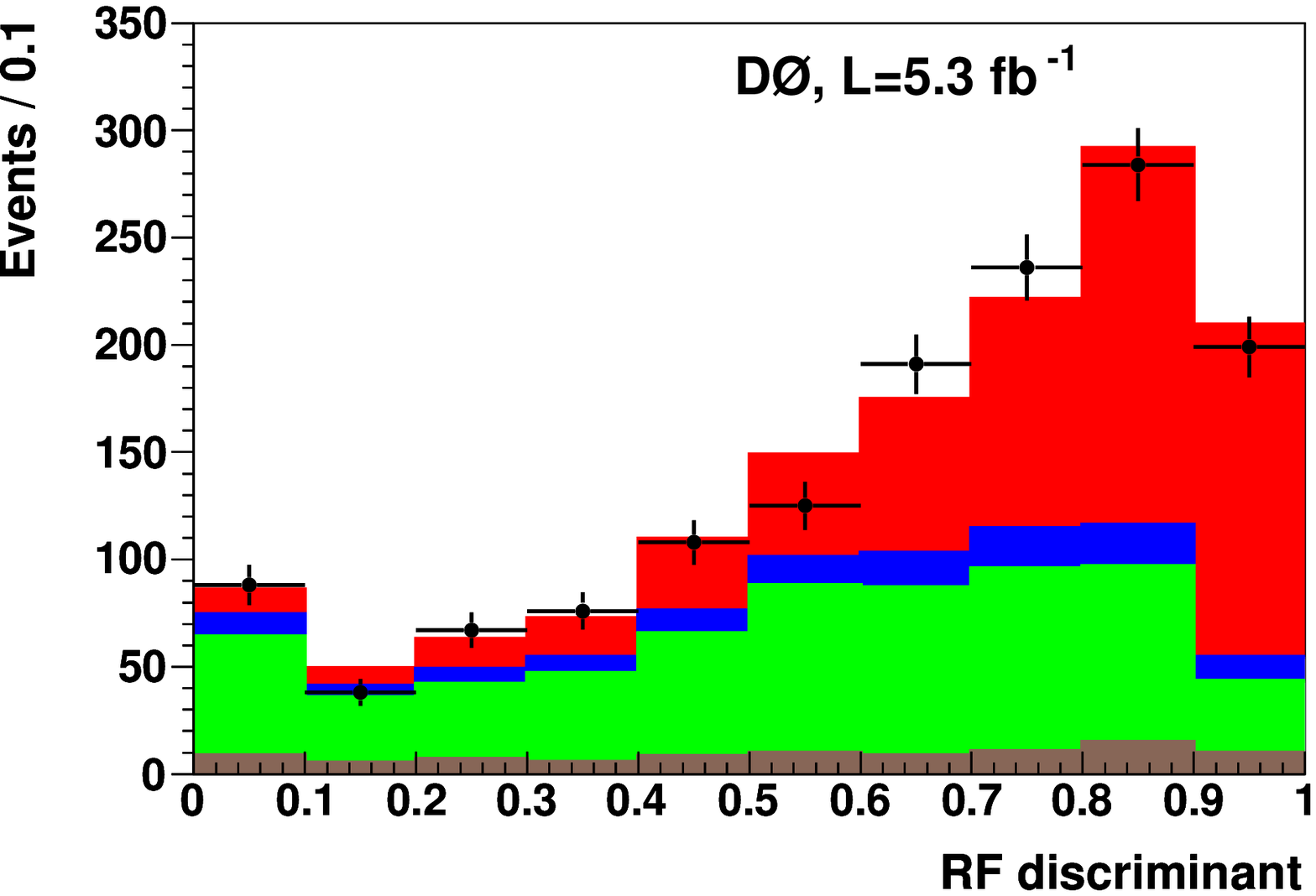}
\vspace{-0.2in}
\caption{Left: Number of data and predicted background events as a function of jet multiplicity, with the number of $\ttbar$ 
t events at the measured cross section events normalized to the measured cross section. The hashed lines represent the uncertainty on the predicted number of events.
Right: Output of the RF discriminant for events with three jets and one $b$-tagged jet for data, backgrounds and 
$\ttbar$ signal normalized to the measured cross section.}
\label{fig:lepjets}
\vspace{-0.2in}
\end{figure}

The inclusive $\ttbar$ production cross section is measured by fitting the $\ttbar$ cross section to data using a binned
maximum likelihood. The likelihood is formed from the data, the $\ttbar$ cross section and the predicted background for 
that cross section. The ``$b$-tagging" method utilizes the event distributions after $b$-jet identification to calculate
the likelihood while the ``kinematics" method constructs a multivariant discriminant to distinguish $\ttbar$ signal 
from background and later uses the discriminant function to obtain the likelihood. The kinematics method exploits the 
kinematic differences between the signal and background before $b$-jet indentification and is therefore not sensitive
to the large systematic uncertainty induced from the $b$-tagging. Using the $b$-tagging method with 4.3 fb$^{-1}$ data, 
CDF experiment measures 
$\sigma_{\ttbar} = 7.22 \pm 0.35\ {\rm (stat)} \pm 0.56\ {\rm (syst)} \pm 0.44\ {\rm (lumi)}\ \rm pb$~\cite{cdf-lepjets-prl}. 
Figure~\ref{fig:lepjets} (left) shows the the predicted number of events for each background process, along with the 
number of expected $\ttbar$ events at the measured cross section compared to data. D0 experiment's $b$-tagging and 
kinematics measurements of $\ttbar$ cross section are described in detail in Ref.~\cite{d0-lepjets-prd}. D0 also uses 
a third method which is a combination of the first two methods: construct a multivariant discriminant (RF) for channels 
dominated by backgrounds, otherwise use $b$-tagging method. The combination takes advantage of the two methods and 
yields a more precise measurement of 
$\sigma_{\ttbar} = 7.78\ ^{+0.77}_{-0.64}\ \rm (stat + syst + lumi)\ pb$~\cite{d0-lepjets-prd} 
using 5.3 fb$^{-1}$ of integrated luminosity for a top quark mass of 172.5~GeV. The discriminant output distributions using 
the combination method for one channel (as an example) is shown in Fig.~\ref{fig:lepjets} (right).
To reduce the large luminosity uncertainty on the $\ttbar$ cross section, CDF measures the $\ttbar$ to 
$Z/\gamma^* \to ll$ ratio in the same corresponding data sample and determine the $\ttbar$ cross section by multiplying 
the ratio by the theoretical $\ttbar$ to $Z/\gamma^* \to ll$ cross section given by the SM. The small uncertainties on the theoretical and 
measured $\ttbar$ to $Z/\gamma^* \to ll$ cross sections are propagated to the final $\ttbar$ cross section measurement.
CDF uses a best linear unbiased estimate (BLUE)~\cite{blue} method to combine the $b$-tagging measurement and 
kinematics measurement and finds 
$\sigma_{\ttbar} = 7.70 \pm 0.52\ \rm pb$~\cite{cdf-lepjets-prl} for $M_t = 172.5$~GeV.

\subsection{Dilepton channel}
In this channel, we require two high-$p_T$ leptons, high missing transverse energy ($\met$) and at least
two jets in the final state. It is independent and orthogonal to the lepton+jets channel and is the only channel
which has a favorable S:B ratio. Two dominant backgrounds are $Z/\gamma^* \to ee/\mu\mu$ with fake $\met$ and $W$+jets
with fake leptons. They are modeled using the data-driven method~\cite{cdf-dilepton-conf}. After event selection,
the final sample contains a high concentration of $\ttbar$ events, which allows us to perform a direct extraction
of $\ttbar$ cross section by $\sigma_{\ttbar} = \frac{N_{obs} - N_{bkg}}{\Sigma_i \mathcal{A}_i \cdot \mathcal{L}_i}$.
$N_{obs}$ is the observed number of dilepton candidate events, $N_{bkg}$ is the total background and $\mathcal{A}_i$
and $\mathcal{L}_i$ are the corrected acceptance and integrated luminosity for analysis channel $i$. CDF measures the
$\ttbar$ cross section using 5.1 fb$^{-1}$ of data for a top mass of 172.5 GeV. The measurement is done before and 
after applying the $b$-tagging requirement and the results are
$\sigma_{\ttbar} = 7.40\ \pm 0.58\ {\rm (stat)}\ \pm 0.63\ {\rm (syst)}\ \pm 0.45\ \rm (lumi)\ pb$ and 
$\sigma_{\ttbar} = 7.25\ \pm 0.66\ {\rm (stat)}\ \pm 0.47\ {\rm (syst)}\ \pm 0.44\ \rm (lumi)\ pb$ correspondingly~\cite{cdf-dilepton-conf}.

\subsection{Tau+jets channel}
Top quark is the heaviest quark and tau is the heaviest lepton, any non-SM mass- or flavor-dependent
couplings could change the top quark decay rate into final states with taus. Therefore it is of interest to measure 
$\sigma(\ppbar \to \ttbar + X) \cdot {\rm BR} (\ttbar \to \tau + jets)$ 
(denoted by ``$\sigma_{\ttbar} \cdot {\rm BR}_{\tau_h + j}$") and compare to the SM prediction.
D0 performs the measurement using semi-hadronic tau decays ($\tau_h$) since secondary electrons and muons from 
tau leptonic decays are difficult to distinguish from primary electrons and muons from $W$ decays. 
The measurement also provides complementary information regarding $\ttbar$ production cross section compared to the
more precise lepton+jets measurements. We select events to have at least four reconstructed 
jets and at least one one $\tau_h$ candidate. In addition, each event must have at least one $b$-jet using
the $b$-tagging algorithm~\cite{btagging}. The main physics backgrounds are the $W$+jets and $Z$+jets contribution
and the main instrumental background is the multijet production. 
We use a neural network ($NN_{sb}$) event discriminant to separate signal
from background and then fit the entire $NN_{sb}$ output distribution to data to extract the numbers of signal and
background events. The measured $\sigma_{\ttbar} \cdot {\rm BR}_{\tau_h + j}$ value is
$0.60\ ^{+0.23}_{-0.22}\ {\rm (stat)}\ ^{+0.15}_{-0.14}\ {\rm (syst)}\ \pm 0.04\ \rm (lumi)\ pb$ for $M_t = 170$~GeV~\cite{d0-tauj-prd}, 
which is consistent with the SM predicted value. 
We repeat the fit while fixing the $\ttbar$ BRs to their SM values and obtain the $\ttbar$ production cross section
$\sigma_{\ttbar} = 6.9 ^{+1.5}_{-1.4}\ \rm pb$~\cite{d0-tauj-prd} for a top quark mass of 170 GeV.  

\subsection{New Physics Searches: $4^{th}$ generation quark $t^{\prime}$}
\begin{figure}[!h!tbp]
\center
\includegraphics[width=0.48\textwidth]{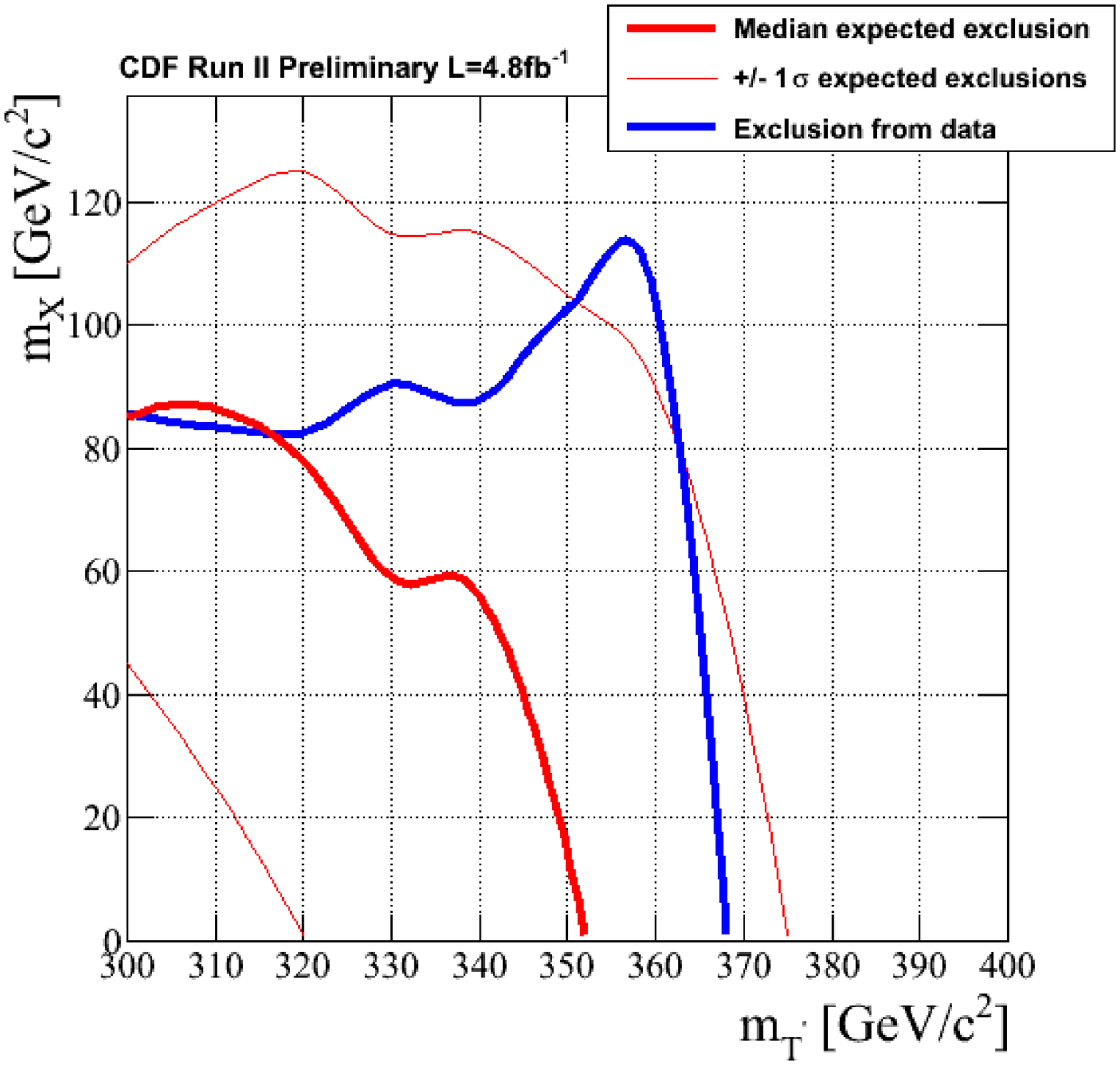}
\includegraphics[width=0.48\textwidth]{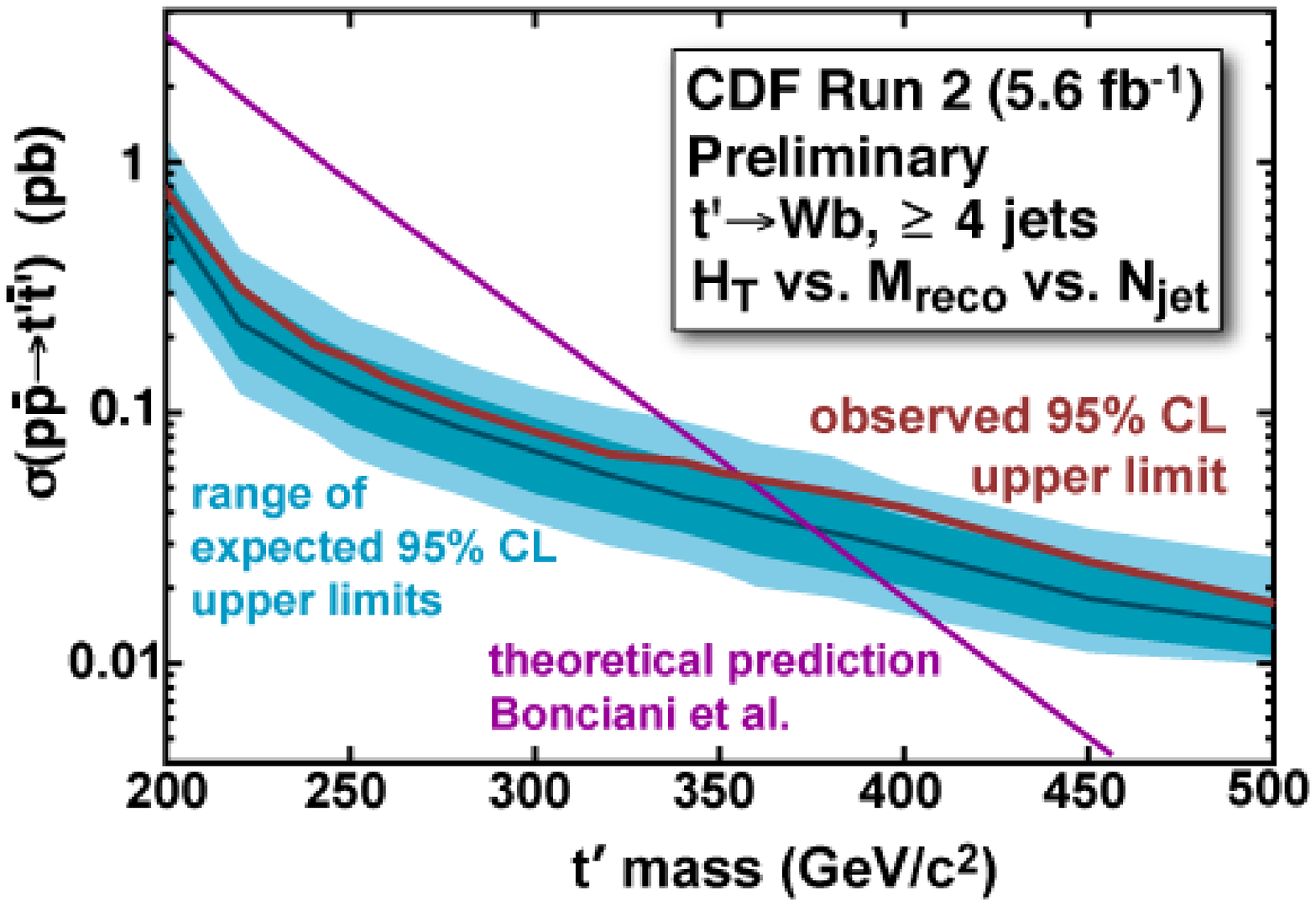}
\vspace{-0.2in}
\caption{Left: Observed and expected exclusion area as a function of ($m_T^{\prime}$, $m_X$).
Right: Observed upper limit at 95\% C.L. on the $t^{\prime}$ production rate as a function of $t^{\prime}$ mass (red curve). 
The purple curve is a theoretical cross section. The blue band represents $\pm 1$ standard deviation expected limit 
(the light blue band corresponds to $\pm 2$ standard deviation).
}
\label{fig:tprime}
\vspace{-0.2in}
\end{figure}

$\ttbar$ production measurements are also useful when searching for new physics, e.g. the $4^{th}$ generation quark 
$t^{\prime}$ search. CDF performs two types of searches for pair production of $t^{\prime}$ using $\ttbar$ event topology. 
One analysis is to search for $t^{\prime}$ decaying via $t^{\prime} \to t + X$, where $X$ is the dark matter particle 
and manifests itself as an excess of missing transverse energy in the detector. The analysis is done in the lepton+jets 
channel with an additional requirement of large $\met$. Another kinematic variable besides $\met$ which is sensitive to 
the signal and background discrimination is the transverse mass of the leptonically decaying $W$ ($mT_W$). The signal 
cross section is extracted by fitting templates of the signal and background shapes in $mT_W$ to the observed number of 
data events taking into account statistical and systematics uncertainties. We obtain the expected and observed upper 
limits on the signal using a Frequentist approach~\cite{cdf-tprime1-conf} done in the two-dimensional (2D) plane of 
($m_T^{\prime}$, $m_X$), where $m_T^{\prime}$ is the mass of the fourth generation quark, and $m_X$ is the mass of the 
dark matter particle. The observed limits are consistent with what the SM predictions.
The final 2D limit is shown in Fig.~\ref{fig:tprime} (left) using 4.8 fb$^{-1}$ of data.

Another search for $t^{\prime}$ is performed in the $Wb$ final state assuming $t^{\prime} \to W + b$. 
We assume that $t^{\prime}$ is produced strongly and has the same couplings as the three generations of the SM quarks.
We use a similar event selection and background modeling as in the $\ttbar$ lepton+jets channel 
measurements. The dominant backgrounds are top pair production and $W$+jet production. 
The new quark is heavier than the top quark and the decay products are more energetic. This effect can be observed
in the total transverse energy variable ($H_T$)~\cite{cdf-tprime2-conf}. It also decays in the same 
chain and allows us to reconstruct its mass in a similar way as in the top mass measurements.
We reconstruct the mass of the $t^{\prime}$ quark ($M_{reco}$) and perform a two-dimensional fit of the
observed ($H_T$,$M_{reco}$) distribution to discriminate the new physics signal from Standard Model
processes. We form a binned likelihood as a function of $t^{\prime}\tbar^{\prime}$ cross section and use a Bayesian 
approach~\cite{cdf-tprime2-conf} to set an upper limit. We generate pseudo-experiments assuming no $t^{\prime}$ signal
and use that to gauge the sensitivity of the analysis. Fig.~\ref{fig:tprime} (right) shows the ranges of the expected
and observed upper limits at 95\% C.L compared to the theoretical calculations. With 5.6 fb$^{-1}$ of data, CDF
excludes the hypothetical $4^{th}$ generation quark $t^{\prime}$ with mass below 358 GeV at 95\% C.L. for $M_t = 172.5$~GeV.

\section{Single Top Quark Production}
Single top quarks are produced via the decay of a time-like virtual $W$ boson accompanied by a bottom quark in 
the s-channel (denoted by ``tb") or via the exchange of a space-like virtual $W$ boson between a light quark and a 
bottom quark in the t-channel (denoted by ``tqb") process.
Previous D0 and CDF publications~\cite{stop-obs-2009-d0,stop-obs-2009-cdf} measured the total single top quark
production cross section assuming the SM predicted ratio between the individual channel's cross sections. 
However several BSM models predict different values of this ratio compared to the SM. Therefore it is of interest
to remove this assumption and measure $s-$channel and $t-$channel production cross section independently.

\begin{figure}[!h!tbp]
\vspace{-0.1in}
\centering
\includegraphics[width=0.48\textwidth]{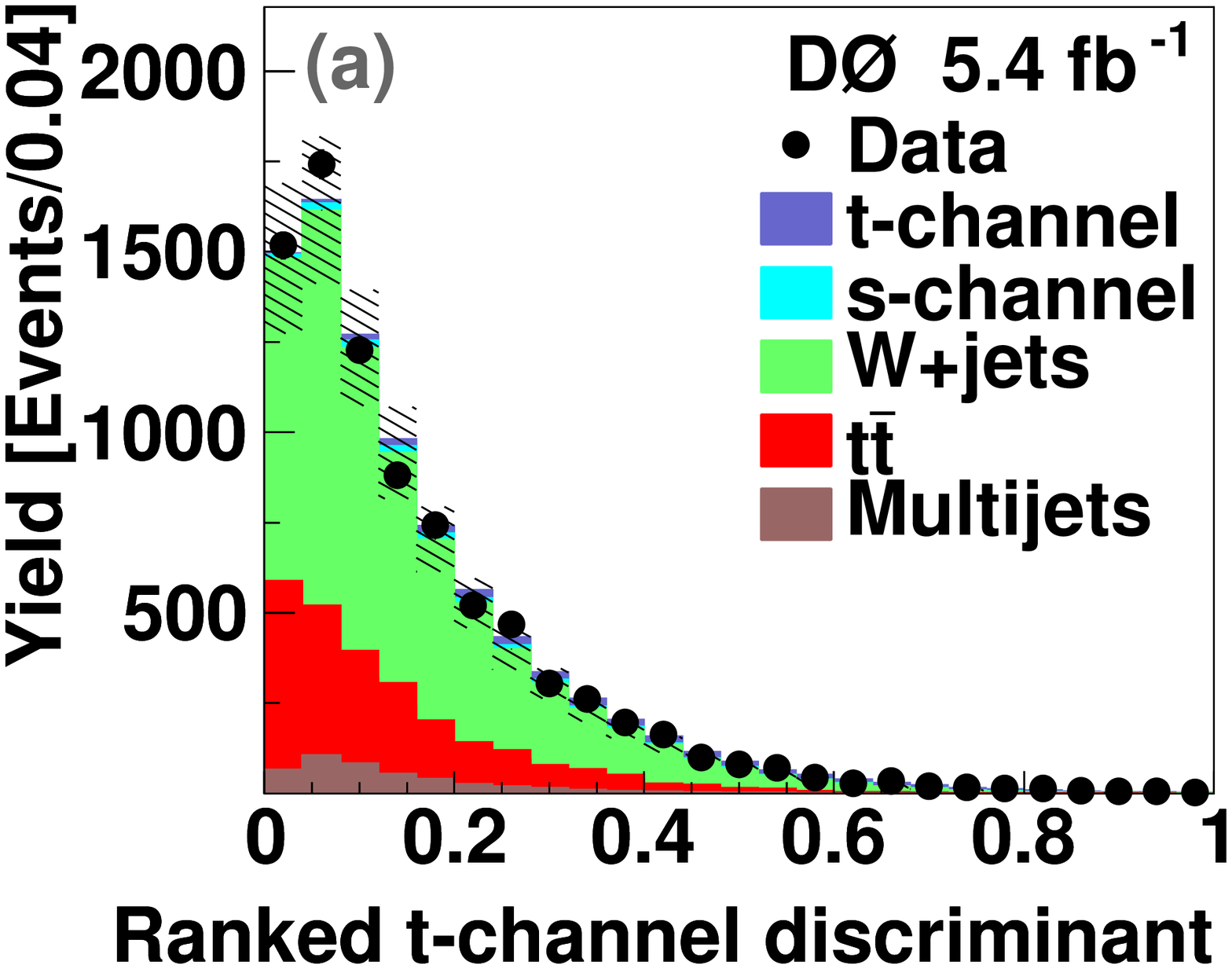}
\includegraphics[width=0.48\textwidth]{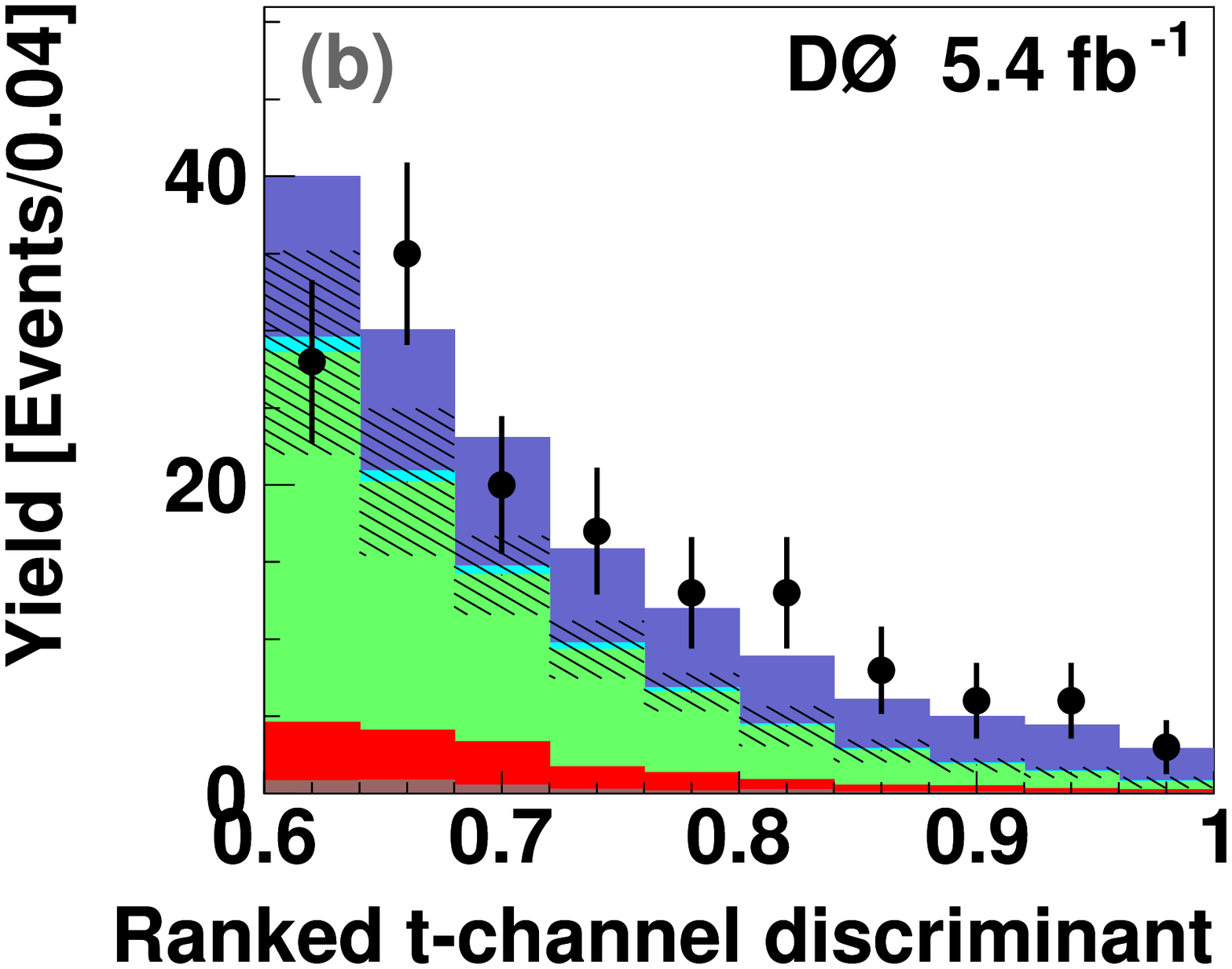}
\vspace{-0.1in}
\caption{Comparison of the signal and background models to data for the combined $t$-channel discriminant
for (a) the entire discriminant range and (b) the signal region.
The bins have been ordered by their expected S:B.
The single top quark contributions are normalized to the measured cross sections. The $t$-channel contribution
is visible above the hatched bands that show the uncertainty on the background prediction.
}
%\vspace{-0.1in}
\label{fig:BNNCombT}
\end{figure}

D0 extends its previous analyses~\cite{stop-obs-2009-d0,d0-prl-2007,d0-prd-2008,t-channel} and performs a new 
measurement on the $t$-channel production rate using a larger dataset of 5.4 fb$^{-1}$ and improved 
techniques~\cite{t-channel-new}.
We require events to have exactly one isolated high-$p_T$ electron or muon, a large $\met$ and two to four 
reconstructed jets (one or two of the jets are identified as $b-$jets~\cite{btagging}). The main backgrounds are
$W$+jets, $\ttbar$ and multijet production.
The largest uncertainties come from the jet energy resolution (JER), corrections to the $b$-tagging efficiency, and
the corrections for the jet-flavor composition in $W$+jets events, with smaller contribution from jet energy scale 
(JES), MC statistics, integrated luminosity, and trigger uncertainties. The total systematic uncertainty on the 
background is 11\%.
We construct multivariant discriminants to improve discrimination between signal and background. We use three methods
to train these discriminants: boosted decision trees (BDT), Bayesian neural
networks (BNN) and neuroevolution of augmented topologies (NEAT). 
We later combine these methods using an additional BNN algorithm that takes
to produces a single combined output discriminant (BNNComb), which further improves the sensitivity and the precision 
of the cross section measurement. Each method is optimized to maximize the sensitivity to the $t-$channel signal
by treating the $s-$channel process as a background component with normalization given by the SM cross
section~\cite{singletop-xsec-kidonakis}. Figure~\ref{fig:BNNCombT} shows comparisons between the $t$-channel signal,
the background model, and data for the combined discriminant,

\begin{figure}[!h!tbp]
\begin{minipage}{0.48\textwidth}
%\vspace{-0.1in}
\centering
\includegraphics[width=\textwidth]{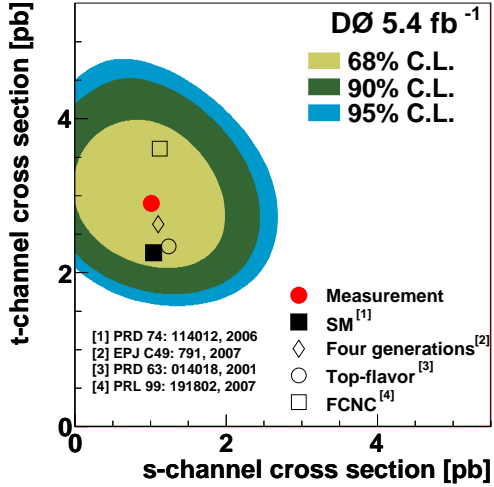}
\vspace{-0.4in}
\caption{Posterior probability density for $t$-channel vs $s$-channel single top quark
production in contours of equal probability density. The measured
cross section and various theoretical predictions are also shown.}
\label{fig:2dposterior}
\end{minipage}
\begin{minipage}{0.48\textwidth}
\begin{center}
\captionof{table}{Measured single top quark production cross sections for different top quark masses.}
\vspace{0.05in}
\begin{tabular}{lccc}
\label{table:BNNcombTMass}
$M_t$   & 170 GeV                & 172.5 GeV              & 175 GeV \\
\hline
$tqb$ & $2.80^{+0.57}_{-0.61}$ & $2.90^{+0.59}_{-0.59}$ & $2.53^{+0.58}_{-0.57}$ \\
$tb$ & $1.31^{+0.77}_{-0.74}$ & $0.98^{+0.62}_{-0.63}$ & $0.65^{+0.51}_{-0.50}$\\
\end{tabular}
\end{center}
%\vspace{-0.1in}
\end{minipage}
\end{figure}

The single top quark production cross section is measured using a Bayesian
approach as in~\cite{d0-prl-2007,d0-prd-2008,stop-obs-2009-d0}.
We follow the approach of \cite{t-channel} and construct a two-dimensional (2D)
posterior probability density as a function of the cross sections for the $s$- and $t$-channel processes.
A binned likelihood is formed using the output discriminants for the signals, backgrounds, and data, 
taking into account all systematic uncertainties and their correlations.
We assume a Poisson distribution for the observed number of data events and nonnegative uniform
prior probabilities for the two cross sections without any assumption on their ratio.
The $t$-channel cross section is then extracted from a one-dimensional posterior
probability density obtained from this 2D posterior
by integrating over the $s$-channel axis, thus not making any assumptions about
the value of the $s$-channel cross section.
Similarly, the $s$-channel cross section is obtained by integrating over the $t$-channel axis.
We generate ensembles of pseudo-experiments to validate the cross section extraction procedure.
Figure~\ref{fig:2dposterior} shows the
2D posterior probability density for the combined discriminant
together with predictions from the SM~\cite{singletop-xsec-kidonakis} and various BSM scenarios~\cite{Alwall:2007,Tait:2000sh,d0-fcnc}.

We measure $\sigma({\ppbar}{\rargap}tqb+X) = 2.90 \pm 0.59$ pb and
$\sigma({\ppbar}{\rargap}tb+X) = 0.98 \pm 0.63$ pb which are in good agreement with the SM predictions for 
a top quark mass of 172.5 GeV~\cite{singletop-xsec-kidonakis}. The significance of the $t$-channel cross section
measurement is computed following a log-likelihood ratio approach~\cite{stop-obs-2009-cdf,t-channel} and is found
to be 5.5 standard deviation (SD) using an asymptotic Gaussian approximation~\cite{fast-significance}.
The measured cross section depends on the assumed mass of the top quark ($M_t$). The dependence is studied
by repeating the analysis on MC samples generated at
different values of $M_t$. Table~\ref{table:BNNcombTMass}
summarizes the measured cross sections for different top quark masses.

%\subsection{Photograph}
%The frame below the author's name is to be replaced by a photograph of the
%author. A paper photo can be sent to the Moriond secretariat along with 
%a hard copy of the manuscript (see below) or a scanned photo can be 
%directly included by replacing:\\
%You may want to include a photograph of yourself below the title
%of your talk. A scanned photo can be 
%directly included by uncommenting (delete the \% sign at the beginning of 
%the lines) the following paragraph in {\bf moriond.sty}:\\
%{\em \%$\backslash$begin\{figure\}[h]}\\
%{\em \%$\backslash$begin\{center\}}\\
%{\em \%$\backslash$psfig\{figure=mypicture.ps,height=35mm\}}\\
%{\em \%$\backslash$end\{center\}}\\
%{\em \%$\backslash$end\{figure\}}\\

\section*{Acknowledgments}
I thank the D0 and CDF collaborations and Fermilab staffs
for producing such nice results on top quark physics. I also would like to 
thank the conference organizers for hosting a wonderful conference with a broad and exciting physics program.
I acknowledge support from DOE (USA).

\end{document}